# Time Delay Estimation in Cognitive Radio Systems

*Invited Paper*

Fatih Kocak*, Hasari Celebi♯, Sinan Gezici*, Khalid A. Qaraqe♯, Huseyin Arslan♮, and H. Vincent Poor†

\* Department of Electrical and Electronics Engineering, Bilkent University, Bilkent, Ankara 06800, Turkey
♯ Department of Electrical and Computer Engineering, Texas A&M University at Qatar, Doha, Qatar
♮ Department of Electrical Engineering, University of South Florida, Tampa, FL, 33620, USA
† Department of Electrical Engineering, Princeton University, Princeton, NJ 08544, USA

*Abstract*— In cognitive radio systems, secondary users can utilize multiple dispersed bands that are not used by primary users. In this paper, time delay estimation of signals that occupy multiple dispersed bands is studied. First, theoretical limits on time delay estimation are reviewed. Then, two-step time delay estimators that provide trade-offs between computational complexity and performance are investigated. In addition, asymptotic optimality properties of the two-step time delay estimators are discussed. Finally, simulation results are presented to explain the theoretical results.

## I. INTRODUCTION

Cognitive radio presents a promising approach to implement intelligent wireless communications systems [1]-[4]. Cognitive radios can be regarded as more capable versions of software defined radios in the sense that they have sensing, awareness, learning, adaptation, goal driven autonomous operation and reconfigurability features [5], [6], which facilitate efficient use of radio resources such as power and bandwidth [1]. As the electromagnetic spectrum is a precious resource, it is important not to waste it. The recent spectrum measurement campaigns in the United States [7] and Europe [8] indicate that the spectrum is under-utilized; hence, opportunistic use of unoccupied frequency bands is highly desirable.

Cognitive radio presents a solution to inefficient spectrum utilization by opportunistically using the available spectrum of a legacy system without interfering with the primary users of that spectrum [2], [3]. In order to facilitate such opportunistic spectrum utilization, cognitive radio devices should be aware of their locations, and monitor the environment continuously. Location awareness requires a cognitive radio device to perform accurate estimation of its position. Cognitive radio devices can obtain position information based on the estimation of position related parameters of signals traveling between them [9], [10]. Among various position related parameters, the time delay parameter provides accurate position information with reasonable complexity [10]. The main focus of this paper is time delay estimation in cognitive radio systems.

The main difference between time delay estimation in cognitive radio systems and that in conventional systems is that a cognitive radio system can transmit and receive over multiple dispersed bands. In other words, as a cognitive radio device can use the spectral holes in a legacy system, it can have a spectrum that consists of multiple bands that are dispersed over a wide range of frequencies (cf. Fig. 1). In [11], the Cramer-Rao lower bounds (CRLBs) for time delay estimation are obtained for dispersed spectrum cognitive radio systems, and the effects of carrier frequency offset (CFO) and modulation schemes of training signals on the accuracy of time delay estimation are quantified. The CRLB expressions imply that frequency diversity can be utilized in time delay estimation. Similarly, the effects of spatial diversity on time delay estimation are investigated in [12] for single-input multiple-output systems. Also, the effects of multiple antennas on time delay estimation and synchronization problems are studied in [13].

This paper studies time delay estimation for dispersed spectrum cognitive radio systems. First, the theoretical limits on time delay estimation are reviewed, and the concept of frequency diversity for time delay estimation is discussed. Then, optimal and suboptimal time delay estimation techniques are studied. Since optimal maximum likelihood (ML) time delay estimation can have very high computational complexity for signals with multiple dispersed bands, two-step time delay estimation techniques are investigated. The proposed two-step time delay estimators first extract unknown parameters related to signals in different frequency bands, and then obtain the final time delay estimate in the second step. In other words, multiple observations (signals at different frequency bands) are processed efficiently to provide a trade-off between computational complexity and estimation performance. In addition, the optimality properties of the two-step estimators are investigated for high signal-to-noise ratios (SNRs), and simulation results are presented to verify the theoretical analysis.

## II. SIGNAL MODEL

Consider a scenario in which $K$ dispersed frequency bands are available to the cognitive radio system, as shown in Fig. 1. The transmitter generates a signal that occupies all the $K$ bands simultaneously, and sends it to the receiver. Then, the receiver is to estimate the time delay of the incoming signal. Since the available bands can be quite dispersed, the use of orthogonal frequency division multiplexing (OFDM) approach [14] can require processing of very large bandwidths. Therefore, processing of the received signal in multiple branches is considered in this study, as in Fig. 2 [11].

For the receiver in Fig. 2, the baseband representation of the received signal in the $i$th branch is given by

$$r_i(t) = \alpha_i \, e^{j\omega_i t} s_i(t-\tau) + n_i(t) \ , \qquad (1)$$

for $i = 1, \ldots, K$, where $\tau$ is the time delay of the signal, $\alpha_i = a_i e^{j\phi_i}$ and $\omega_i$ represent, respectively, the channel coefficient and the CFO for the signal in the $i$th branch, $s_i(t)$ is the baseband representation of the transmitted signal in the $i$th band, and $n_i(t)$ is complex white Gaussian noise with independent components, each having spectral density $\sigma_i^2$. It is assumed that the signal in each branch can be modeled as a narrowband signal; hence, a single complex channel coefficient is used to represent the fading of each signal. Also, it should be noted that the effects of CFO are considered in the signal model in (1) since multiple down-conversion units are employed in the receiver, as shown in Fig. 2.

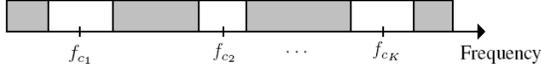

Fig. 1

ILLUSTRATION OF DISPERSED SPECTRUM UTILIZATION IN THE COGNITIVE RADIO SYSTEM, WHERE THE WHITE SPACES REPRESENT THE AVAILABLE BANDS.

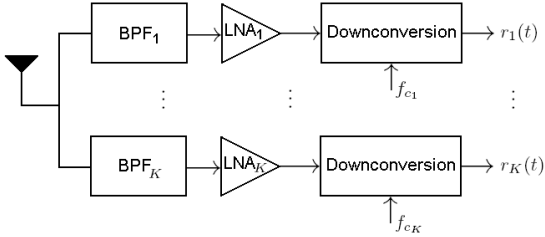

Fig. 2

BLOCK DIAGRAM OF THE FRONT-END OF THE COGNITIVE RADIO RECEIVER.

## III. THEORETICAL LIMITS ON TIME DELAY ESTIMATION

Estimation of the time delay parameter $\tau$ based on $K$ received signals in (1) involves $(3K+1)$ unknown parameters since the channel coefficients and CFOs are also unknown. In other words, the vector of unknown parameters, $\boldsymbol{\theta}$, can be expressed as $\boldsymbol{\theta} = [\tau \; a_1 \cdots a_K \; \phi_1 \cdots \phi_K \; \omega_1 \cdots \omega_K]$.

For an observation interval of $[0, T]$, the log-likelihood function for $\boldsymbol{\theta}$ is expressed as[1] [15]

$$\Lambda(\boldsymbol{\theta}) = c - \sum_{i=1}^{K} \frac{1}{2\sigma_i^2} \int_0^T \left| r_i(t) - \alpha_i e^{j\omega_i t} s_i(t-\tau) \right|^2 dt, \quad (2)$$

where $c$ is a constant that is independent of $\boldsymbol{\theta}$. Then, the Fisher information matrix (FIM) [15] can be obtained from (2) as in [11], and the inverse of the FIM can be used to obtain the CRLB on mean-squared errors (MSEs) of unbiased time delay estimators. In its most generic form, the CRLB can be expressed as [11]

$$E\{(\hat\tau - \tau)^2\} \geq \left( \sum_{i=1}^{K} \frac{a_i^2}{\sigma_i^2} \left( \tilde{E}_i - (\hat{E}_i^{\mathrm{R}})^2 / E_i \right) - \xi \right)^{-1}, \quad (3)$$

where $E_i = \int_0^T |s_i(t-\tau)|^2 dt$ is the signal energy, $\tilde{E}_i = \int_0^T |s_i'(t-\tau)|^2 dt$, with $s'(t)$ representing the first derivative of $s(t)$, and $\hat{E}_i^{\mathrm{R}} = \int_0^T \mathcal{R}\{s_i'(t-\tau_i) s_i^*(t-\tau_i)\} dt$, with $\mathcal{R}$ denoting the operator that selects the real-part of its argument. In addition, $\xi$ represents a term that depends on the spectral properties of signals $s_i(t)$ for $i = 1, \ldots, K$ [11].

The CRLB expression in (3) reveals that the accuracy of time delay estimation depends on the SNR at each branch (via the $a_i^2/\sigma_i^2$ terms), as well as on the properties of signals $s_i(t)$ in (1). In addition, the summation term in (3) indicates that the accuracy can be improved as more bands are employed, which implies that frequency diversity can be utilized in time delay estimation. For instance, when one of the bands is in a deep fade (that is, small $a_i^2$), some other bands can still be in

[1]The unknown parameters are assumed to be constant for $t \in [0, T]$.

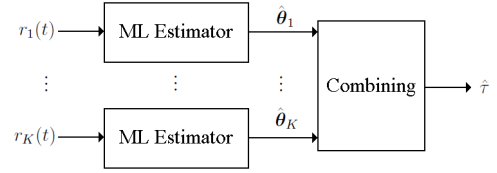

Fig. 3

THE BLOCK DIAGRAM OF THE TWO-STEP TIME DELAY ESTIMATION APPROACH. THE SIGNALS $r_1(t), \ldots, r_K(t)$ ARE AS SHOWN IN FIG. 2.

good condition to facilitate accurate time delay estimation.

In order to investigate the effects of signal design on the time delay estimation accuracy, suppose that the baseband representation of the signals in different branches are of the form $s_i(t) = \sum_l d_{i,l} p_i(t - lT_i)$, where $d_{i,l}$ denotes the complex training data and $p_i(t)$ is a pulse with duration $T_i$. Then, the $\xi$ term in (3) becomes zero, which results in a CRLB expression that would be obtained in the absence of CFOs [11]. In other words, the effects of CFOs can be mitigated via appropriate signal design.

In the special case of $|d_{i,l}| = |d_i| \; \forall l$ and $p_i(t)$ satisfying $p_i(0) = p_i(T_i)$ for $i = 1, \ldots, K$, (3) reduces to [11]

$$E\{(\hat\tau - \tau)^2\} \geq \left( \sum_{i=1}^{K} \frac{\tilde{E}_i \, a_i^2}{\sigma_i^2} \right)^{-1}. \quad (4)$$

Hence, for linearly modulated signals with constant envelopes, improved time delay estimation accuracy can be achieved.

## IV. TWO-STEP TIME DELAY ESTIMATION

The ML estimate of $\boldsymbol{\theta}$ can be obtained from (2) as

$$\hat{\boldsymbol{\theta}}_{\mathrm{ML}} = \arg\max_{\boldsymbol{\theta}} \sum_{i=1}^{K} \frac{1}{\sigma_i^2} \int_0^T \mathcal{R}\left\{ \alpha_i^* e^{-j\omega_i t} r_i(t) s_i^*(t-\tau) \right\} dt - \frac{E_i a_i^2}{2\sigma_i^2} \quad (5)$$

which requires an optimization over a $(3K+1)$-dimensional space, hence is quite impractical in general. Therefore, a two-step time delay estimation approach is considered in this study, as shown in Fig. 3. In the first step, each branch of the receiver performs estimation of the time delay, the channel coefficient and the CFO related to the signal in that branch. Then, in the second step, the estimates from all the branches are used to obtain the final time delay estimate.

### A. First Step: Parameter Estimation at Different Branches

In the first step, the unknown parameters of each received signal are estimated at the corresponding receiver branch according to the ML criterion (cf. Fig. 3). Based on the signal model in (1), the log-likelihood function at branch $i$ becomes

$$\Lambda_i(\boldsymbol{\theta}_i) = c_i - \frac{1}{2\sigma_i^2} \int_0^T \left| r_i(t) - \alpha_i e^{j\omega_i t} s_i(t-\tau) \right|^2 dt, \quad (6)$$

for $i = 1, \ldots, K$, where $\boldsymbol{\theta}_i = [\tau \; a_i \; \phi_i \; \omega_i]$ denotes the vector of unknown parameters related to the signal at the $i$th branch, $r_i(t)$, and $c_i$ is a constant that is independent of $\boldsymbol{\theta}_i$.

From (6), the ML estimator at branch $i$ is expressed as

$$\hat{\boldsymbol{\theta}}_i = \arg\min_{\boldsymbol{\theta}_i} \int_0^T \left| r_i(t) - \alpha_i e^{j\omega_i t} s_i(t-\tau) \right|^2 dt, \quad (7)$$

where $\hat{\boldsymbol{\theta}}_i = [\hat{\tau}_i \ \hat{a}_i \ \hat{\phi}_i \ \hat{\omega}_i]$ is the vector of estimates at the $i$th branch. After some manipulation, (7) yields

$$\left[\hat{\tau}_i \ \hat{\phi}_i \ \hat{\omega}_i\right] = \arg\max_{\phi_i,\omega_i,\tau_i} \left| \int_0^T \mathcal{R}\left\{r_i(t)\,e^{-j(\omega_i t+\phi_i)} s_i^*(t-\tau_i)\right\} dt \right| \tag{8}$$

$$\hat{a}_i = \frac{1}{E_i} \int_0^T \mathcal{R}\left\{r_i(t)\,e^{-j(\hat{\omega}_i t+\hat{\phi}_i)} s_i^*(t-\hat{\tau}_i)\right\} dt . \tag{9}$$

In other words, at each branch, optimization over a three-dimensional space is performed to obtain the unknown parameters, which is significantly less complex than the ML estimation in (5) that requires optimization over $(3K+1)$ variables.

### B. Second Step: Combining Estimates from Different Branches

After obtaining time delay estimates $\hat{\tau}_1, \ldots, \hat{\tau}_K$ in (8), the second step combines those estimates according to one of the criteria below and makes the final time delay estimate.

*1) Optimal Combining:* According to the "optimal" combining[2] criterion, the time delay estimate is obtained as

$$\hat{\tau} = \frac{\sum_{i=1}^{K} \kappa_i \hat{\tau}_i}{\sum_{i=1}^{K} \kappa_i} , \tag{10}$$

where $\hat{\tau}_i$ is the time delay estimate of the $i$th branch, which is obtained from (8), and $\kappa_i = \hat{a}_i^2 \tilde{E}_i / \sigma_i^2$. In other words, the optimal combining approach estimates the time delay as a weighted average of the time delays at different branches, where the weights are chosen as proportional to the multiplication of the SNR estimate, $E_i \hat{a}_i^2 / \sigma_i^2$, and $\tilde{E}_i / E_i$. As $\tilde{E}_i$ is defined as the energy of the first derivative of $s_i(t)$, $\tilde{E}_i / E_i$ can be expressed, using Parseval's relation, as $\tilde{E}_i / E_i = 4\pi^2 \beta_i^2$, where $\beta_i$ is the effective bandwidth of $s_i(t)$, which is defined as $\beta_i^2 = \frac{1}{E_i} \int_{-\infty}^{\infty} f^2 |S_i(f)|^2 df$ with $S_i(f)$ denoting the Fourier transform of $s_i(t)$ [15]. Therefore, it is observed that the optimal combining technique assigns a weight to the time delay estimate of a given branch in proportion to the product of the SNR estimate and the effective bandwidth related to that branch. The intuition behind this combining approach is that signals with larger effective bandwidths and/or larger SNRs facilitate more accurate time delay estimation [15]; hence, their weights are larger in the combining process. This intuition will be verified in Section IV-C theoretically.

*2) Selection Combining (SC):* Another approach to obtain the final time delay estimate is to determine the "best" branch and to use its estimate as the final time delay estimate. According to SC, the best branch is defined as the one with the maximum value of $\kappa_i = \hat{a}_i^2 \tilde{E}_i / \sigma_i^2$ for $i = 1, \ldots, K$. That is, the branch with the maximum multiplication of the SNR estimate and the effective bandwidth is selected as the best branch and its estimate is used as the final one. In other words,

$$\hat{\tau} = \hat{\tau}_m , \qquad m = \arg\max_{i\in\{1,\ldots,K\}} \left\{\hat{a}_i^2 \tilde{E}_i / \sigma_i^2\right\} , \tag{11}$$

where $\hat{\tau}_m$ represents the time delay estimate at the $m$th branch.

*3) Equal Combining:* The equal combining approach assigns equal weights to the estimates from different branches and obtains the time delay estimate as $\hat{\tau} = \frac{1}{K}\sum_{i=1}^{K} \hat{\tau}_i$.

Considering the combining techniques above, it is observed that they are counterparts of the diversity combining techniques employed in communications systems [16]. However, the main distinction is that the aim is to maximize the SNR or to reduce the probability of symbol error in communications systems [16], whereas, in the current problem, it is to reduce the MSE of the time delay estimation. In other words, this study focuses on diversity combining for time delay estimation, where the diversity is due to the dispersed spectrum utilization of the cognitive radio system.

### C. Optimality Properties of Two-Step Time Delay Estimation

In this section, the asymptotic optimality properties of the two-step time delay estimators are investigated in the absence of CFO. In order to analyze the performance of the estimators at high SNRs, the result in [12] for time-delay estimation at multiple receive antennas is considered first.

**Lemma 1 [12]:** *Assume that $\int_{-\infty}^{\infty} s_i'(t-\tau) s_i^*(t-\tau) dt = 0$ for $i = 1, \ldots, K$. Then, for the signal model in (1), the delay estimate in (8) and the channel amplitude estimate in (9) can be modeled, at high SNR, as*

$$\hat{\tau}_i = \tau + \nu_i \quad and \quad \hat{a}_i = a_i + \eta_i , \tag{12}$$

*for $i = 1, \ldots, K$, where $\nu_i$ and $\eta_i$ are independent zero mean Gaussian random variables with variances $\sigma_i^2/(\tilde{E}_i\,a_i^2)$ and $\sigma_i^2/E_i$, respectively. In addition, $\nu_i$ and $\nu_j$ ($\eta_i$ and $\eta_j$) are independent for $i \neq j$.*

From Lemma 1, it is obtained that $\mathrm{E}\{\hat{\tau}_i\} = \tau$ for $i = 1, \ldots, K$. In other words, the time delay estimates of all branches are asymptotically unbiased. Since the combining techniques in the previous section considers only one, or a linear combinations of the time delay estimates at different branches, the two-step time delay estimation techniques have an asymptotic unbiasedness property.

Regarding the variance of the estimators, it is first shown that the optimal combining technique has a variance that is approximately equal to the CRLB at high SNRs.[3] To that aim, the conditional variance of $\hat{\tau}$ in (10) given $\hat{a}_1, \ldots, \hat{a}_K$ is expressed as follows:

$$\mathrm{Var}\{\hat{\tau}|\hat{a}_1,\ldots,\hat{a}_K\} = \frac{\sum_{i=1}^{K} \kappa_i^2 \mathrm{Var}\{\hat{\tau}_i|\hat{a}_1,\ldots,\hat{a}_K\}}{\left(\sum_{i=1}^{K}\kappa_i\right)^2} , \tag{13}$$

where the independence of the time delay estimates is used to obtain the result (cf. Lemma 1). Since $\mathrm{Var}\{\hat{\tau}_i|\hat{a}_1,\ldots,\hat{a}_K\} = \mathrm{Var}\{\hat{\tau}_i|\hat{a}_i\} = \sigma_i^2/(\tilde{E}_i\,a_i^2)$ from Lemma 1 and $\kappa_i = \hat{a}_i^2 \tilde{E}_i/\sigma_i^2$, (13) can be manipulated to obtain

$$\mathrm{Var}\{\hat{\tau}|\hat{a}_1,\ldots,\hat{a}_K\} = \sum_{i=1}^{K} \frac{\hat{a}_i^4 \tilde{E}_i}{a_i^2 \sigma_i^2} \left(\sum_{i=1}^{K} \frac{\hat{a}_i^2 \tilde{E}_i}{\sigma_i^2}\right)^{-2} . \tag{14}$$

Lemma 1 states that $\hat{a}_i$ is distributed as a Gaussian random variable with mean $a_i$ and variance $\sigma_i^2/E_i$ at high SNRs. Hence, for sufficiently large values of $\frac{E_1}{\sigma_1^2}, \ldots, \frac{E_K}{\sigma_K^2}$, (14) can be approximated by $\left(\sum_{i=1}^{K} \frac{\tilde{E}_i\,a_i^2}{\sigma_i^2}\right)^{-1}$, which is equal to the CRLB expression in (4). Therefore, the optimal combining technique in (10) yields an approximately optimal estimator at high SNRs.

---

[2] The optimality property is investigated in Section IV-C.

[3] This is the main reason why this combining technique is called *optimal*.

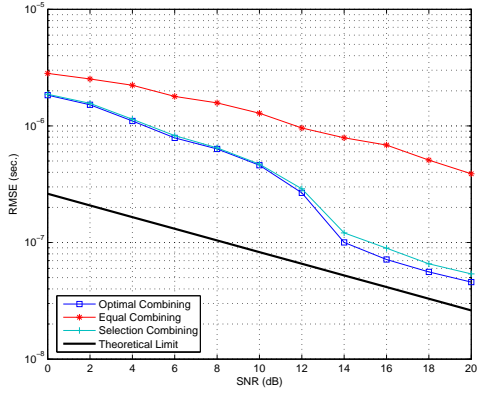

Fig. 4

RMSE VS. SNR FOR THE TWO-STEP ALGORITHMS, AND THE THEORETICAL LIMIT (CRLB). THE SIGNAL OCCUPIES THREE DISPERSED BANDS WITH $B_1 = 200$ KHZ, $B_2 = 100$ KHZ AND $B_3 = 400$ KHZ.

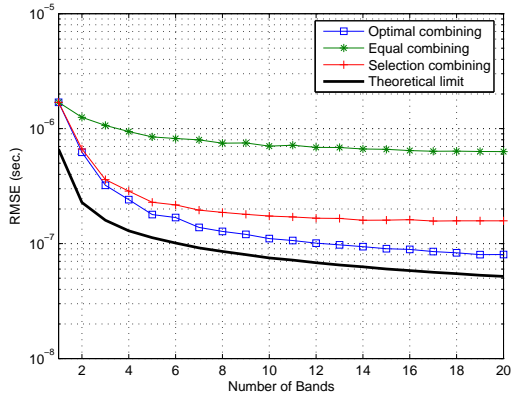

Fig. 5

RMSE VS. THE NUMBER OF BANDS FOR THE TWO-STEP ALGORITHMS, AND THE THEORETICAL LIMIT (CRLB). EACH BAND IS 100 KHZ WIDE, AND $\sigma_i^2 = 0.1 \ \forall i$.

Regarding the selection combining approach in (11), the conditional variance can be approximated at high SNR as

$$\mathrm{Var}\{\hat{\tau}|\hat{a}_1,\ldots,\hat{a}_K\} \approx \min\left\{\frac{\sigma_1^2}{\tilde{E}_1 a_1^2},\ldots,\frac{\sigma_K^2}{\tilde{E}_K a_K^2}\right\}. \quad (15)$$

In general, the SC approach performs worse than the optimal combining technique. However, when the estimate of a branch is significantly more accurate than the others, its performance can get very close to that of the optimal combining technique.

Finally, for the equal combining technique in Section IV-B.3, the variance can be calculated as $\mathrm{Var}\{\hat{\tau}\} = \frac{1}{K^2}\sum_{i=1}^{K}\frac{\sigma_i^2}{\tilde{E}_i a_i^2}$. The equal combining approach is expected to have the worst performance since it does not make use of any information about the SNR or the signal bandwidths in the estimation of the time delay, as investigated next.

## V. SIMULATION RESULTS AND CONCLUSIONS

In this section, simulations are performed to evaluate the CRLBs and the performance of the time delay estimators. Signal $s_i(t)$ in (1) at branch $i$ is modeled by a unit-energy Gaussian doublet as in [11] with bandwidth $B_i$. In all the simulations, the spectral densities of the noise at different branches are assumed to be equal; that is, $\sigma_i = \sigma$ for $i = 1,\ldots,K$. Also, the SNR of the system is defined with respect to the total energy of the signals at different branches, i.e., $\mathrm{SNR} = 10\log_{10}\left(\frac{\sum_{i=1}^{K} E_i}{2\sigma^2}\right)$.

In assessing the root-mean-squared errors (RMSEs) of the different estimators, a Rayleigh fading channel is considered. Namely, the channel coefficient $\alpha_i = a_i\,\mathrm{e}^{j\phi_i}$ in (1) is modeled as $a_i$ being a Rayleigh distributed random variable and $\phi_i$ being uniformly distributed over $[0, 2\pi)$. In addition, the same average power is assumed for all the bands; that is, $\mathrm{E}\{|\alpha_i|^2\} = 1$ is used. The time delay, $\tau$, in (1) is uniformly distributed over the observation interval, and it is assumed that there is no CFO in the system.

First, the performance of the two-step estimators is evaluated with respect to the SNR for a system with $K = 3$, $B_1 = 200$ kHz, $B_2 = 100$ kHz and $B_3 = 400$ kHz. The results in Fig. 4 indicate that the optimal combining technique has the best performance as expected from the theoretical analysis, and SC, which estimates the delay according to (11), has performance close to that of the optimal combining technique. On the other hand, the equal combining technique has significantly worse performance than the others, as it combines all the delay estimates equally. Since the delay estimates of some branches can have very large errors due to fading, the RMSEs of equal combining become quite significant. Finally, it is observed that the performance of the optimal combining technique gets quite close to the CRLB at high SNRs, in agreement with the asymptotic arguments in Section IV-C.

Next, the RMSEs of the two-step estimators are plotted against the number of bands in Fig. 5, where each band is assumed to have 100 kHz bandwidth. In addition, the spectral densities are set to $\sigma_i^2 = \sigma^2 = 0.1 \ \forall i$. From Fig. 5, it is observed that the optimal combining has better performance than the selection combining and equal combining techniques. Also, as the number of bands increases, the amount of reduction in the RMSE per additional band decreases (i.e., diminishing return). In fact, the selection combining technique seems to converge to a constant value for large numbers of bands. This is intuitive as the selection combining technique always uses the estimate from one of the branches; hence, in the presence of a sufficiently large number of bands, additional bands do not result in a significant increase in the diversity. On the other hand, the optimal combining technique has a slope that is quite similar to that of the CRLB; that is, it makes use of the frequency diversity efficiently.

## REFERENCES

[1] J. Mitola and G. Q. Maguire, "Cognitive radio: Making software radios more personal," *IEEE Personal Commun. Mag.*, vol. 6, no. 4, pp. 13–18, Aug. 1999.
[2] S. Haykin, "Cognitive radio: Brain-empowered wireless communications," *IEEE J. Select Areas Commun.*, vol. 23, no. 2, pp. 201–220, Feb. 2005.
[3] Z. Quan, S. Cui, H. V. Poor, and A. H. Sayed, "Collaborative wideband sensing for cognitive radios," *IEEE Signal Process. Mag.*, vol. 25, no. 6, pp. 60–73, Nov. 2008.
[4] Q. Zhao and B. Sadler, "A survey of dynamic spectrum access," *IEEE Signal Process. Mag.*, vol. 24, no. 3, pp. 79–89, May 2007.
[5] J. O. Neel, "Analysis and design of cognitive radio networks and distributed radio resource management algorithms," Ph.D. dissertation, Virginia Polytechnic Inst. and State Univ., Blacksburg, VA, Sep. 2006.
[6] H. Celebi and H. Arslan, "Enabling location and environment awareness in cognitive radios," *Elsevier Computer Communications*, vol. 31, no. 6, pp. 1114–1125, April 2008.
[7] Federal Communications Commission (FCC), "Facilitating opportunities for flexible, efficient, and reliable spectrum use employing cognitive radio technologies," *ET Docket No. 03-108*, Mar. 2005.

[8] M. Wellens, J. Wu, and P. Mahonen, "Evaluation of spectrum occupancy in indoor and outdoor scenario in the context of cognitive radio," in *Proc. International Conference on Cognitive Radio Oriented Wireless Networks and Communications*, Orlando, FL, Aug. 2007.
[9] H. Celebi and H. Arslan, "Cognitive positioning systems," *IEEE Trans. Wireless Commun.*, vol. 6, no. 12, pp. 4475–4483, Dec. 2007.
[10] S. Gezici, "A survey on wireless position estimation," *Wireless Personal Communications, Special Issue on Towards Global and Seamless Personal Navigation*, vol. 44, no. 3, pp. 263–282, Feb. 2008.
[11] S. Gezici, H. Celebi, H. V. Poor, and H. Arslan, "Fundamental limits on time delay estimation in dispersed spectrum cognitive radio systems," *IEEE Trans. Wireless Commun.*, vol. 8, no. 1, pp. 78–83, Jan. 2009.
[12] S. Gezici and Z. Sahinoglu, "Ranging in a single-input multiple-output (SIMO) system," *IEEE Commun. Lett.*, vol. 12, pp. 197–199, Mar. 2008.
[13] C. Williams, S. McLaughlin, and M. A. Beach, "Exploiting multiple antennas for synchronization," *IEEE Trans. Vehicular Technology*, vol. 58, no. 2, pp. 773–787, Feb. 2009.
[14] T. A. Weiss and F. K. Jondral, "Spectrum pooling: An innovative strategy for the enhancement of spectrum efficiency," *IEEE Commun. Mag.*, vol. 42, no. 3, pp. 8–14, March 2004.
[15] H. V. Poor, *An Introduction to Signal Detection and Estimation*. New York: Springer-Verlag, 1994.
[16] A. Goldsmith, *Wireless Communications*. Cambridge, UK: Cambridge University Press, 2005.